# Remagnetization of bulk high-temperature superconductors subjected to crossed and rotating magnetic fields

P Vanderbemden[1], Z Hong[2], T A Coombs[2], M Ausloos[3], N Hari Babu[4], D A Cardwell[4] and A M Campbell[4]

(1) SUPRATECS and Department of Electrical Engineering and Computer Science B28, Sart-Tilman, B-4000 Liège, Belgium

(2) Centre for Advanced Photonics and Electronics, Engineering Department, University of Cambridge, 9 JJ Thomson Avenue, Cambridge CB3 0FA, United Kingdom

(3) SUPRATECS and Department of Physics B5, Sart-Tilman, B-4000 Liège, Belgium

(4) IRC in Superconductivity, University of Cambridge, Madingley Road, Cambridge CB3 OHE, United Kingdom

E-mail : Philippe.Vanderbemden@ulg.ac.be

**Abstract.** Bulk melt-processed Y-Ba-Cu-O (YBCO) has significant potential for a variety of high field permanent magnet-like applications, such as the rotor of a brushless motor. When used in rotating devices of this kind, however, the YBCO can be subjected to both transient and alternating magnetic fields that are not parallel to the direction of magnetization and which have a detrimental effect on the trapped field. These effects may lead to a long-term decay of the magnetization of the bulk sample. In the present work, we analyze both experimentally and numerically the remagnetization process of a melt-processed YBCO single domain that has been partially demagnetized by a magnetic field applied orthogonal to the initial direction of trapped flux. Magnetic torque measurements are used as a tool to probe changes in the remanent magnetization during various sequences of applied field. The application of a small magnetic field between the transverse cycles parallel to the direction of original magnetization results in partial remagnetization of the sample. Rotating the applied field, however, is found to be much more efficient at remagnetizing the bulk material than applying a magnetizing field pulse of the same amplitude. The principal features of the experimental data can be reproduced qualitatively using a two-dimensional finite-element numerical model based on an *E-J* power law. Finally, the remagnetization process is shown to result from the complex modification of current distribution within the cross-section of the bulk sample.

# 1. Introduction

Single domain bulk melt-processed RE-Ba-Cu-O ((RE)BCO, where RE represents a Rare Earth ion) has significant potential for use in permanent magnet-like engineering applications [1], such as magnetic bearings [2,3] and high power density rotating machines [4-8] due to its ability to trap relatively large magnetic fields [9,10]. In the so-called "trapped-field" or "pre-fluxed" synchronous motor, for example, the bulk (RE)BCO artefact is placed in the machine rotor and permanently magnetized parallel to its *c*-axis before the machine is started [4,5]. In such a configuration, the superconductor, which acts effectively as a permanent magnet, is expected to follow continuously the rotating magnetic field produced by the three-phase stator windings. In practice, however, any sudden variation of the applied torque on the shaft may cause the magnetization of the sample to mis-align with the direction of the stator field for a short period of time [8]. In addition, the sample itself may effectively be exposed to a rotating field during such a process due to the reaction with the armature reaction and to inhomogeneities of the stator field itself [11]. Such phenomena result generally in the reduction of the field trapped in the superconductor, with the most severe demagnetizing effects arising when the AC field perturbations are perpendicular to the direction of permanent magnetization [12].

The situation in which a magnetized bulk HTS sample is subjected to a field applied orthogonal to the original direction of magnetization (i.e. the so-called "*crossed magnetic fields*" configuration) has been studied extensively for several years [13-18]. It is now well established that small transverse field cycles may cause significant decay of the sample magnetization, which is also known as "*collapse of magnetic moment*" [14,15]. The repetition of a transverse field cycle in rotating machinery, therefore, may result in a long-term decay of the field produced by any bulk HTS magnet used in the design of the device and result, ultimately, in its failure. Unlike in a standard "crossed field" experiment where the sample is de-magnetized by a series of *transverse* field cycles *only*, a superconducting permanent magnet used in a machine rotor is also subjected to *parallel* fields that can re-magnetize the sample. Detailed understanding of the remagnetization process in this configuration remains unknown. One relevant issue, however, is to determine the minimum amplitude of applied magnetic field required to recover fully the original magnetization following the demagnetization process. It is also necessary to determine whether the field is *rotated continuously* from the transverse to the parallel direction by the application of successive pulses of transverse and parallel field.

Several theoretical approaches have been made in the literature to attempt to describe the magnetic behaviour of irreversible type-II superconductors subjected to crossed and rotating magnetic fields, including the double critical-state model [19,20], the two-velocity hydrodynamic model [14,15], the elliptic flux-line-cutting critical-state model [21-23] and a number of variational approaches [24-27]. Although some of these models can be applied to samples of finite size, the studies performed to date do not take demagnetization effects into account (i.e. the length of the sample is assumed infinite in the direction of the applied fields, such as the case of an infinite slab with crossed magnetic fields parallel to the slab surface). In contrast, different theoretical approaches have been developed in the case of thin strips [28,29] or very thin platelets (thickness $d \ll$ dimensions $w$, $L$) subjected to a DC field applied perpendicular to the platelet face with an AC field applied perpendicular to the platelet thickness [30]. Strictly speaking, none of these approaches applies to the case of the bulk HTS magnet geometry used in most engineering applications, which predominantly require short, disc-shaped samples characterized by an aspect ratio (thickness:lateral dimension of the large face) of typically 1 : 3.

We reported previously the use of a Finite Element Method (FEM) [31] to model the two-dimensional magnetization of superconductors of finite thickness in the crossed field configuration [32-35]. Rather than assuming a vertical $E$-$J$ characteristic, which is the approach of the techniques described above, our method involves modelling the superconductor with a $E$-$J$ power law relation, $E \propto J^n$, with $n \gg 1$ [36,37]. In addition, the currents are assumed to flow perpendicularly to the sample cross-section and too close at infinity in our particular geometry. As a result, the electric field, $E$, and the current density, $J$, are maintained parallel to each other and are always orthogonal to the magnetic field $H$. An important consequence of this geometry is that it avoids both flux free configurations and flux cutting, which simplifies the physical understanding of the problem. Strikingly, this simple approach was able to reproduce successfully the main features of the collapse of magnetic moment under the action of a transverse field [31].

The objective of the present work is to investigate both experimentally and numerically the conditions of re-magnetization of the bulk sample after application of a transverse field. This involves measuring the magnetization of a YBCO single domain under crossed and rotating magnetic fields by torque magnetometry and comparing the results with those of a two-dimensional model in order to analyze the physical processes involved in remagnetizing the sample and to draw conclusions of practical significance.

## 2. Experiment

Bulk melt-processed single domains of YBCO, consisting of a superconducting $YBa_2Cu_3O_{7-\delta}$ (Y-123) matrix with discrete $Y_2BaCuO_5$ (Y-211) inclusions, were fabricated by a top seeded melt growth (TSMG) technique, as described in refs. [38-41]. Small samples with an aspect ratio of ~ 1:3 (typical size 0.8 x 0.8 x 0.25 mm) and faces coinciding with the major crystallographic planes of the Y-123 unit cell were cut from the parent single domain using a wire saw. These samples were characterized magnetically for applied field conditions of $H \parallel ab$ and $H \parallel c$ at $T = 77$ K using a Quantum Design SQUID Magnetic Property Measurement System (MPMS). The properties of one of these samples are summarized in Table 1.

Measurements of the YBCO samples in the crossed-field configuration were carried out using magnetic torque magnetometry in a Quantum Design Physical Property Measurement System (PPMS). In all experiments, the melt-processed sample was initially magnetized parallel to the $c$-axis by field cooling (FC) down to $T = 77$ K in an applied field of 0.5 T. The field amplitude was intentionally much larger than the full-penetration field of the sample ($\mu_0 H_p = 74$ mT) to ensure that the maximum (trapped) magnetic moment $m_c$ is achieved [42,43]. A constant time interval of approximately 5 minutes was allowed for magnetic relaxation following removal of the applied (*c*-axis) field [43-45]. A transverse magnetic field $H_{ab}$ was then applied parallel to the *ab* plane of the sample and the magnetic torque τ recorded, as illustrated schematically in Fig. 1(a). In this case, the magnetic torque can be expressed as;

$$\underline{\tau} = \underline{m}_c \times \mu_0 \underline{H}_{ab} \quad (1)$$

The variation of magnetic moment $m_c$ *vs.* the applied field $H_{ab}$ can be determined from the measured torque by $m_c(H_{ab}) = \tau(H_{ab})/(\mu_0 H_{ab})$, since $m_c$ and $H_{ab}$ are mutually orthogonal. This technique is particularly convenient for determining accurately the magnetic moment of small-size superconductors subjected to crossed and rotating fields. One constraint of the method, however, is that the measurements must always be performed in the presence of a small but finite transverse field $H_{ab}$. As shown in Fig. 1(b), the measured $m_c(H_{ab})$ data obtained by torque measurements agree perfectly with the initial value of magnetic moment determined by SQUID magnetometry in the absence of a transverse field. The collapse in magnetization shown in Fig. 1(b) is also qualitatively similar to that reported in the literature [13-17] and to Hall probe measurements performed on larger samples of the same material [18,31].

Three different sequences of applied field were performed and the results compared in order to analyze the remagnetization process, as illustrated schematically in Fig. 2(a). In the classical demagnetization sequence ("S-trans"), the transverse field is cycled several times between 0 and $+H_{max}$ and the magnetic moment measured at $H = +H_{max}$, as indicated by the arrows in Fig. 2. A transverse field cycle is again employed in the so-called "remagnetization" sequence ("S-rem"), but with the incorporation of an intermediate field cycle stage between each transverse field cycle in which magnetic field is applied parallel to the direction of original magnetization. In this case, the amplitude of the remagnetizing field ($H_{max}$) is identical to that of the transverse field and the magnetic moment is measured at each transverse field maxima. Finally, in the so-called "S-rot" sequence, the applied transverse field is ramped initially to $H = +H_{max}$ and rotated repeatedly by 90° to the parallel direction ($H||c$) and back to the transverse direction ($H||ab$). In this case, the magnetic moment is measured when the field is perpendicular to the magnetic moment. Experimentally, the three sequences were adjusted so that the time interval between two successive measurements is constant (500 s).

A similar set of applied field sequences was also performed, but under conditions of double polarity, as shown in Fig. 2(b). This involved (i) the "D-trans" sequence (cycling the transverse field between $+H_{max}$ and $-H_{max}$), (ii) the "D-rem" sequence (cycling the applied magnetic field parallel to the direction of original magnetization once between the transverse field cycles) and (iii) the "D-rot" sequence (rotating the field by 180° from the $+H_{max}$ to the $-H_{max}$ directions during each transverse field cycle).

In addition, a separate rotation sequence was investigated where a field of constant amplitude $H_{max}$ was applied parallel to the direction of trapped flux, rotated at some angle $\alpha$ and then cycled between two symmetric angular positions $+\alpha$ and $-\alpha$. The particular case of $\alpha = 90°$ corresponds to the "rotation" sequence in the condition of double polarity described above. In this experiment, the magnetic torque $\tau$ was measured when the applied field was directed at angle $+\alpha$ with respect to the $c$-axis, in which case the magnetic moment $m_c$ is given by $\tau / (\mu_0 H_{max} \sin \alpha)$.

## 3. Modelling parameters

The numerical method used for modelling the electromagnetic behaviour of a bulk superconductor is based on solving the set of Maxwell equations in two dimensions using the Finite-Element Method (FEM) software *Comsol Multiphysics 3.2* and a numerical evaluation scheme described in detail in ref. [32]. The *E-J* behaviour of the superconducting material is modelled assuming $\underline{E} = E_c (\underline{J}/J_c)^n$, where $E_c$ (= 1 μV/cm) is the threshold electric field used to define the critical current density $J_c$. A constant $n = 21$ value was used, corresponding to typical values of this parameter reported for bulk melt-processed YBCO [46,47]. From the SQUID magnetization measurement results, we used a field dependent $J_c(B) = J_{c0} (1 + |B|/B_0)^{-1}$ [48], where $|B|$ denotes the modulus of the local magnetic induction. The constant parameters $J_{c0}$ and $B_0$ were determined to reproduce the experimental $M(H)$ curve for $H\|c$; the procedure yields $J_{c0} = 10^5$ A/cm$^2$ and $B_0 = 16$ mT.

Space is assumed to be infinite in the $x$ direction (i.e. perpendicular to the plane of the paper) in the two-dimensional model used in this study, as shown in Fig. 1(c). The direction of the remanent magnetization and transverse applied fields are denoted hereafter by $z$ and $y$. The sample is assumed to consist of an infinite cylinder of rectangular cross-section $y_0 \times z_0$. In this geometry, the magnetic flux lies in the $y$-$z$ plane and the current density field lines flow in the $x$ direction. Although the two-dimensional geometry used in the model does not represent the true geometry of the superconducting samples used in the experiments, the key point is that the sample has *finite* dimensions along both the remanent magnetization ($z$) and the applied transverse field ($y$) directions. In the present model, the sample was magnetized initially by a field pulse of duration 0.1 s applied parallel to $z$. A constant time interval (10 s) was then employed in order to allow the trapped magnetic moment to relax due to flux creep effects. The sample was then subjected to the applied magnetic field sequences described in section 2 and the program was used to compute both the average magnetic moment along the $z$ direction and the current distribution $J(x)$ within the sample cross-section.

## 4. Results and discussion

*4.1. Magnetization sequences in the conditions of single polarity*

Figure 3(a) compares the evolution of the measured *c*-axis magnetic moment $m_c$ with the number of sweeps *N* for the three unipolar magnetic field sequences ("S-trans, S-rem and S-rot") shown in Fig. 2(a). The amplitude of magnetic field $H_{max}$ is half that of the full-penetration field $H_p$ parallel to the *ab* plane. It should be remembered that the full-penetration fields determined for $H\|c$ and for $H\|ab$ differ by ~15% in the case of the sample studied here (c.f. table 1). In the following discussion, the term "penetration field $H_p$" will always refer to the value determined for a field configuration of $H\|ab$, i.e. $\mu_0 H_p = 64$ mT. It is apparent from Fig. 3(a) that the (*N*=1) data points in each sequence are very close to each other since they result from the application of the same $0.5\,H_p$ transverse field. A subsequent series of transverse field cycles only ("S-trans") leads to a monotonic decrease of the magnetization, as is observed usually [14, 17, 18]. Applying a magnetizing field pulse parallel to the initial direction of magnetization between each transverse field cycle ("S-rem") yields a slightly less pronounced decay in magnetization. Rotating the applied field by 90° to the parallel direction and back to the transverse direction ("S-rot") is found to be more efficient in remagnetizing the sample than applying a magnetizing field pulse ($H\|c$) of the same amplitude.

Figure 3(b) shows the results of the model, corresponding to the same experimental conditions as those described above. From a quantitative point of view, the normalized magnetization decays resulting from the various field cycles are found to be less pronounced than those measured experimentally. From a qualitative point of view, however, the modelled $m_z(N)$ curves are remarkably similar to the experimental data. Rotating the applied field after the first transverse field sweep increases the magnetic moment ($m_Z(N=2) > m_Z(N=1)$ for the "S-rot" curve), whereas further rotations lead to a slow decay of the magnetization. The rotation mode always remagnetizes the sample more efficiently than that obtained by incorporating a parallel field cycle stage between the transverse field cycles.

In order to identify the possible reasons of such behaviour, we have modelled the current distribution within the sample cross-section at several selected times between the two first data points ($N = 1$ and $N = 2$) for each of the 3 sequences. The results of this model are presented in Fig. 4. The initial current distribution that gives rise to a positive trapped magnetic moment (i.e. $m_z > 0$) is shown in Fig. 4(a). The two half cross-sections carry currents with opposite signs , although the magnitude of current density is not uniform within each half of the cross-section; this feature arises because a field-dependent $J_c(B)$ is used in the model [31]. The application of a positive transverse field $H_y = 0.5\,H_p$ leads to a reversal of the current density in the top-left and bottom-right regions of the sample, as shown in Fig. 4(b); a thin horizontal layer located at the top (resp. the bottom) of the sample carries negative current (resp. positive). These current directions correspond to those required for shielding the increasing applied field $dH_y/dt > 0$. Consequently, both upper and lower layers of the sample no longer contribute efficiently to the *z*-axis magnetic moment $m_z$. This corresponds to the (*N*=1) data point in Fig. 3.

Both the "S-trans" and "S-rem" sequences involve decreasing the transverse field to zero (Fig. 4(c)) and yield identical results. In both cases, the shielding currents in the top and bottom layers change their sign, as required for shielding the decreasing applied field $dH_y/dt < 0$. In the rotation mode ("S-rot"), the decreasing transverse field is coupled to an increasing parallel field that induces negative current (resp. positive) at the left (resp. right)-hand side of the sample, as required for shielding the increasing applied field $dH_z/dt > 0$. The corresponding current distribution is shown in Fig. 4(c). Fig. 4(d) shows the current distribution after application of the magnetizing pulse in the "S-rem" mode. In this case, two "double layers" containing currents of opposite signs have appeared at the bottom-left and the top-right of the sample. The current distribution corresponding to the second data point ($N = 2$) is shown in Fig. 4(e). In the "S-trans" and "S-rem" modes, the application of a second increasing transverse field $dH_y/dt > 0$ has imposed the sign of the current density at the top and at the bottom of the sample. The current distributions in these modes are very similar to each other, resulting in very close values of magnetic moment $m_z$. Conversely, a rotation of the magnetic field from the parallel to the transverse direction ("S-rot" mode) has imposed the sign of the current density along the 4 faces of the sample, as required for shielding both the increasing transverse field $dH_y/dt > 0$ *and* the decreasing applied field $dH_z/dt < 0$. The resulting magnetic moment $m_z$ is noticeably larger than that of the two other modes (cf. Fig. 3).

These results can be summarized schematically by considering the sign of current density in the outer zones of the sample cross-section, as illustrated in Fig. 5. The left (resp. right) part initially contains positive (resp. negative) currents, giving rise to a positive trapped magnetic moment $m_z > 0$. When transverse or parallel magnetic fields are applied, the current distribution is modified according to the sign of the electric field induced by magnetic flux density variations $dB/dt$ in various regions of the sample. The increasing transverse field $H_y$ tries to impose the sign of the current density at the top and bottom parts of the sample; $J < 0$ at the top, $J > 0$ at the bottom. Similarly, the parallel field $H_z$ tries to impose the sign of the current density along the left-hand and right-hand edges of the sample. In the present case, we focus on the current distribution that remains after the parallel (remagnetizing) field has decreased to 0, i.e. $J > 0$ at the left, $J < 0$ at the right of the sample. A conflict occurs at two corners of the sample (the top-left and the bottom-right) where the positive and the remagnetizing fields try to impose current of opposite signs. The results of the model shown in Fig. 4 suggest the following simple rules for determining the sign of the current density in these regions: (i) the field applied most recently later imposes its sign if the fields are applied sequentially; (ii) the field corresponding to the largest $dB/dt$ imposes its sign if the fields are applied simultaneously. When the transverse field is applied *after* the remagnetizing field, the two pairs of corners that are symmetric with the $z$-axis carry currents of the same sign and no longer contribute to the $z$-axis magnetic moment. When the fields are applied *simultaneously* (rotation of the magnetic field), however, the 4 corners recover currents that have the same sign as the initial current distribution and can therefore contribute efficiently to the $z$-axis magnetization. The assumption of relatively simple physical phenomena involved in the magnetization process, illustrated schematically in Fig. 5, account remarkably well for the efficiency of the "S-rot" magnetization mode compared to the "S-rem" mode, as is both observed experimentally and predicted by the model.

*4.2. Influence of the field amplitude in the rotation mode*

We focus now on rotating repeatedly the field by 90° between the transverse and parallel directions ("S-rot" mode). Fig. 6 shows the variation of the measured *c*-axis magnetic moment $m_c$ with the number of rotations *N* for several amplitudes of magnetic field ranging from 0.125 $H_p$ to 2 $H_p$. In each case, the magnetic moment is normalized with respect to its initial value. The (*N*=1) measurements refer to the remaining magnetic moment after the first application of the transverse field; the corresponding values decrease as the field amplitude increases, as expected [18]. In all cases, the first magnetic field rotation increases the magnetic moment: $m_c(N=2) > m_c(N=1)$. At low amplitudes ($H \leq 0.75 H_p$), further rotations yield a monotonic decay of the magnetization whereas at large amplitudes ($H > 0.75 H_p$), the magnetic moment stabilizes at a plateau of maxima for field amplitudes approximately equal to $H_p$ (inset of Fig. 6). The existence of an optimum applied field can be understood as follows. At small amplitudes of rotation, the parallel field is not large enough to remagnetize the sample. At large amplitudes of rotation, the sample is almost fully remagnetized between the cycles but the large value of the transverse field is responsible for a significant decay of the *c*-axis magnetic moment. This decay is ascribed to the modification of the current distribution within the sample as well as a reduction of the current amplitude at every point of the sample because of the $J_c(B)$ dependence. A field amplitude of $H_p$ was found to be sufficient to provide a satisfactory remagnetization of the sample in the experimental conditions employed in this study.

*4.3. Magnetization sequences in the conditions of double polarity*

The measured magnetic moment $m_c$ *vs*. number of sweeps *N* curves for the three bipolar magnetic field sequences ("D-trans, D-rem and D-rot") are compared in Fig. 7(a) for an amplitude of magnetic field equal to 0.5 $H_p$. In comparison with the results obtained in the unipolar sequences (Fig. 3), the corresponding decays are significantly larger although the qualitative features are very similar. In particular, rotating the applied field by 180° between the two transverse field directions ("D-rot" sequence) is again more efficient at remagnetizing the sample than incorporating a remagnetizing field pulse ($H||c$) between the transverse cycles. The modelled current distribution (not shown here) suggests that the sign of induced currents at the corners of the sample cross-section is responsible of such a behaviour, as is the case for the sequences under single polarity (Fig. 5). The modelled magnetic moment (Fig. 7(b)) under identical experimental conditions is fully consistent with the experimental measurements of magnetic moment.

Finally, we investigated the rotation of a field of constant amplitude $H_{max}$ cycled between two symmetric angular positions +α and -α with respect to the sample *c*-axis. Two different cycles are compared in Fig. 8: (i) $H_{max} = 0.5 H_p$, α = 90° and (ii) $H_{max} = H_p$, α = 30°. Both of these cycles share a common characteristic in that the amplitude of the field component perpendicular to the initial direction of magnetic moment ($m_c$) is equal to 0.5 $H_p$ (inset of Fig. 8). The magnetic moment $m_c$ is measured when the transverse field component is maximum and equal to 0.5 $H_p$. As may be seen in Fig. 8, the magnetic moments measured at $H=0.5 H_p$, $\alpha = 90°$ are significantly larger than those measured at $H=H_p$, $\alpha = 30°$. After a few (~ 8) rotation cycles, the normalized magnetic moment appears to stabilize at saturation levels of 0.78 ($H=0.5 H_p$, $\alpha = 90°$) and 0.25 ($H=H_p$, $\alpha = 30°$). This can be understood by modelling the current distribution within the sample cross-section in each case (Fig. 9). The initial current distribution leading to a permanent magnetic moment is shown in Fig. 9(a). The application of a field parallel to the *z*-axis direction induces shielding currents that oppose the initial current distribution. The regions where these currents circulate is larger for larger for $H=H_p$ than for $H=0.5 H_p$, as shown in Fig. 9(b). Rotating the magnetic field distorts noticeably the current distribution (Fig. 9(c)). When the field is oriented at 30°, the *z*-component (equal to $H_p \cos(30°) \sim 0.87 H_p$) is associated with shielding currents that oppose the field and, therefore, reduce the magnetic moment significantly, as experimentally observed ("*N*=1" point in Fig. 8). The current distributions at the end of the second sweep (i.e. corresponding to the "*N*=2" points in Fig. 8) are shown in Fig. 9(d). For $H=0.5 H_p$ and $\alpha = 90°$, both right-hand and left-hand border sides of the sample carry currents that have the same sign as the initial current distribution and contribute efficiently to a positive $m_z$. Conversely, for $H=H_p$ and $\alpha = 30°$, the two border zones carrying currents giving rise to a positive $m_z$ are restricted to a small region near the horizontal middle plane of the sample, which explains reasonably the small value of the magnetic moment in this configuration. Further field rotations were modelled in order to obtain the current distribution after 10 cycles (Fig. 9(e)). The corresponding modelled magnetic moments (70 % and 6% of the initial value) are again consistent with the experimental data.

In order to investigate the detrimental effect of a given number (*N*=10) of field rotations on the trapped field on the initial moment, the model was used to compute the remaining current distribution when the magnetic field is rotated back to the initial direction of magnetic moment and decreased to 0 after the last cycle. As can be seen in Fig. 9(f), the current distributions display an intricate "yin-yang" shape resulting from the preceding applied field cycles. The associated magnetic moments are 73% and 122% of the initial magnetic moment, respectively. The existence of a final magnetic moment that is *larger* than the initial value can be explained by remembering that the initial current distribution shown in Fig. 9(a) is obtained after a constant time interval (~ 10 s) has elapsed following initial magnetization; the magnetic moment decreases by approximately ~ 25-30% during this period [31]. If the same time interval is allowed for magnetic relaxation after the last cycle, *both* remaining moments are smaller than the initial value, as expected intuitively. The values of the remaining magnetic moments calculated after the applied field has been switched off cannot be compared to the experimental data because the torque measurements reported here must be performed in the presence of a finite transverse field. Nevertheless, the results of the model suggest clearly that the effects of tilting periodically the applied field around the direction of the permanent magnetic moment while maintaining a constant amplitude of the transverse field component are much less severe for small tilt angles. In addition, the decay of the magnetic moment resulting from such rotation cycles is much less pronounced than the "magnetization collapse" measured with transverse fields only (cf. Fig. 7).

## 5. Conclusions

We have analyzed both experimentally and numerically the remagnetization process of a melt-processed YBCO single domain that has been partially demagnetized by a magnetic field applied orthogonal to the initial direction of trapped flux. Magnetic torque measurements have proved an efficient tool for determining accurately the magnetic moment of small-size superconductors subjected to crossed and rotating fields. The experimental data could be reproduced successfully using a two-dimensional finite-element model based on a *E-J* power law and avoiding both flux free and flux cutting configurations. The value of magnetization was shown to be governed predominantly by the sign of the current density at the corners of the sample cross-section. This feature cannot be predicted by a model that assumes an infinite length in the direction of the crossed and rotating fields. In the present experimental conditions (aspect ratio ~ 1:3, field-dependent $J_c$), rotating the applied field by 90° or 180° between the two transverse field directions was shown to be more efficient at remagnetizing the sample than inserting a remagnetizing field pulse ($H$||c) of the same amplitude between the transverse cycles. A field amplitude on the order of the full-penetration field was found to be sufficient to provide a satisfactory remagnetization of the sample. The comparison between two "tilting field" configurations corresponding to the same amplitude of the transverse field component showed that the smaller rotation angle leads to a much less pronounced decay of the magnetic moment. The differences in the values of remaining magnetic moments after the different transverse / rotation cycles were shown to result from the complex distribution of the induced currents within the sample cross-section.

**Acknowledgments**

Ph.V. is grateful to the FNRS for a travel grant. We acknowledge J. F. Fagnard, Profs. R. Cloots and B. Vanderheyden for fruitful discussions and comments. We also thank the FNRS, the ULg and the Royal Military Academy (RMA) of Belgium for cryofluid and equipment grants.

**Table**

Table 1. Full-penetration field and remanent magnetic moment determined for the YBCO sample at $T$ = 77 K. The sample dimensions are 0.855 x 0.855 x 0.26 mm, the shortest side is parallel to the $c$-axis.

| Field direction | Full penetration field $\mu_0 H_p$ | Remanent magnetic moment |
|---|---|---|
| H ∥ *ab* plane | 64 mT | 2.7 $10^{-6}$ Am² |
| H ∥ *c*-axis | 74 mT | 8.9 $10^{-6}$ Am² |

**Figure captions**

**Figure 1**. (a) Schematic illustration of the sample mounted on the platform used for torque measurements. (b) Magnetic moment ∥ *c*-axis during the application of one sweep of transverse field ∥ *ab*. The closed symbol indicates the initial trapped magnetic moment without transverse field. The open symbols represent data determined from magnetic torque measurements. (c) Geometry used for the two-dimensional model: the sample is of cross-section $y_0 \times z_0$ and is infinite in the $x$ direction. In all cases, the sample is pre-magnetized along the $z$-axis and the transverse field is applied parallel to the $y$-axis.

**Figure 2**. Sequences of magnetic fields applied to the sample. The transverse field has either (a) single or (b) double polarity.

**Figure 3**. (a) Measured magnetic moment ∥ *c*-axis at the end of each unipolar magnetic field cycle shown in Fig. 2. The magnetic moment is normalized with respect to its initial value $m_0$ determined by SQUID measurements and the field amplitude is 0.5 $H_p$. (b) Modelled magnetic moment under the same experimental conditions.

**Figure 4**. Modelled data of the current density distribution $J_x(y, z)$ within the cross-section of the sample at selected times during each of the three cycles of magnetic field ("S-trans, S-rem and S-rot") of amplitude $H_{max}$ = 0.5 $H_p$. (a) Initial state; (b) transverse field increased to $H_{max}$; (c) transverse field component reduced to 0; (d) state after a magnetizing field pulse is applied in the "S-rem" cycle; (e) transverse field component increased to $H_{max}$.

**Figure 5**. Schematic illustration of the sample cross-section during the application of demagnetizing and remagnetizing fields.

**Figure 6**. Measured magnetic moment || $c$-axis at the end of each rotation cycle ("S-rot") for several amplitudes of magnetic field ranging from 0.125 $H_p$ to 2 $H_p$. The magnetic moment is normalized with respect to its initial value $m_0$. Inset : measured magnetic moment || $c$-axis at the end of the 6$^{th}$ cycle as a function of the normalized amplitude of magnetic field $H / H_p$.

**Figure 7**. (a) Measured magnetic moment || $c$-axis at the end of each bipolar magnetic field cycle shown in Fig. 2. The magnetic moment is normalized with respect to its initial value $m_0$; the field amplitude is 0.5 $H_p$. (b) Modelled magnetic moment using the same experimental conditions.

**Figure 8**. Normalized magnetic moment $m_c$ measured at the end of each cycle involving the rotation of a magnetic field $H_{max}$ between two symmetric angular positions +α and -α with respect to $c$-axis. Inset: schematic representation of the cycles ($H_{max}$ = 0.5 $H_p$, α = 90° and $H_{max}$ = $H_p$, α = 30°); the magnetic moment is measured when the transverse field component is maximum (black arrow).

**Figure 9**. Modelled data of the current density distribution $J_x(y, z)$ within the cross-section of the sample at selected times during two cycles involving the rotation of a magnetic field $H_{max}$ between two symmetric angular positions +α and -α with respect to the initial direction of magnetic moment (||$z$). (a) Initial state; (b) field applied parallel to $z$-direction; (c) initial rotation of the field at an angle +α ; (d) field rotated a second time; (e) state after the 10$^{th}$ rotation cycle ; (f) state after the field is rotated back parallel to the z-axis and then decreased to 0.

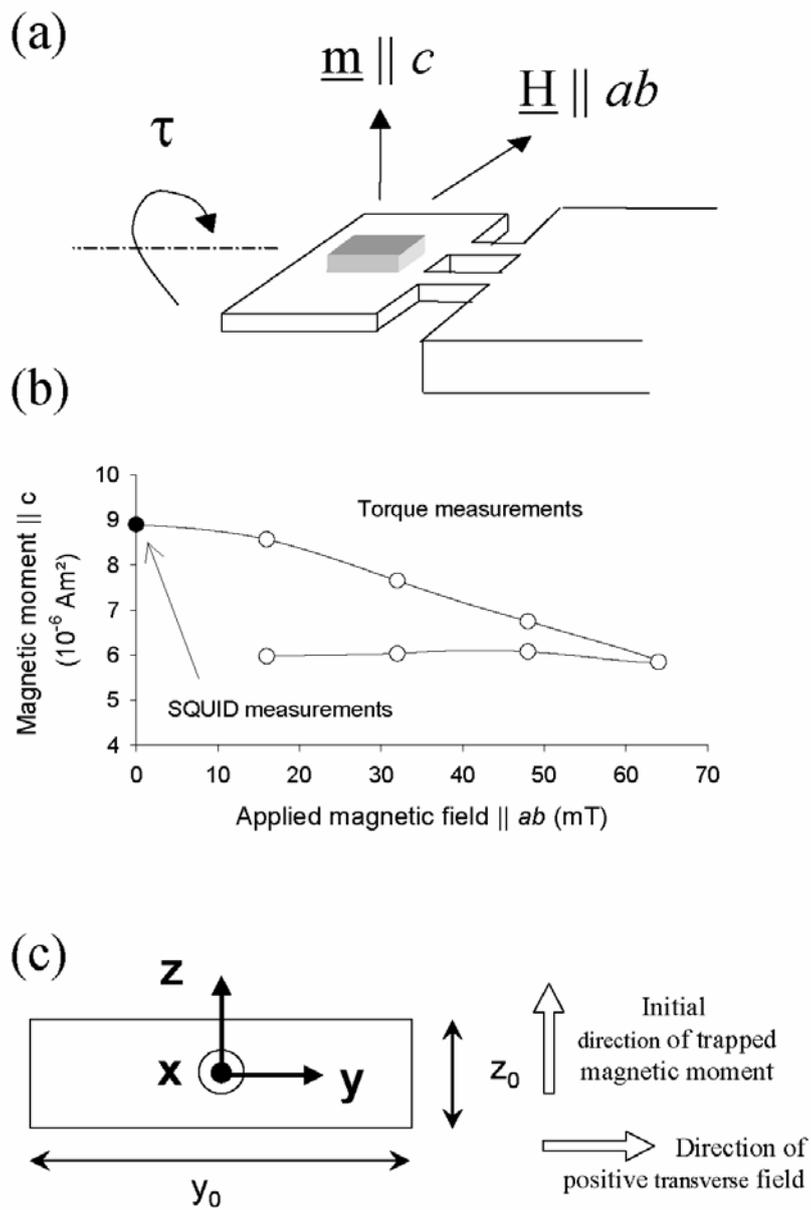

**Figure 1**

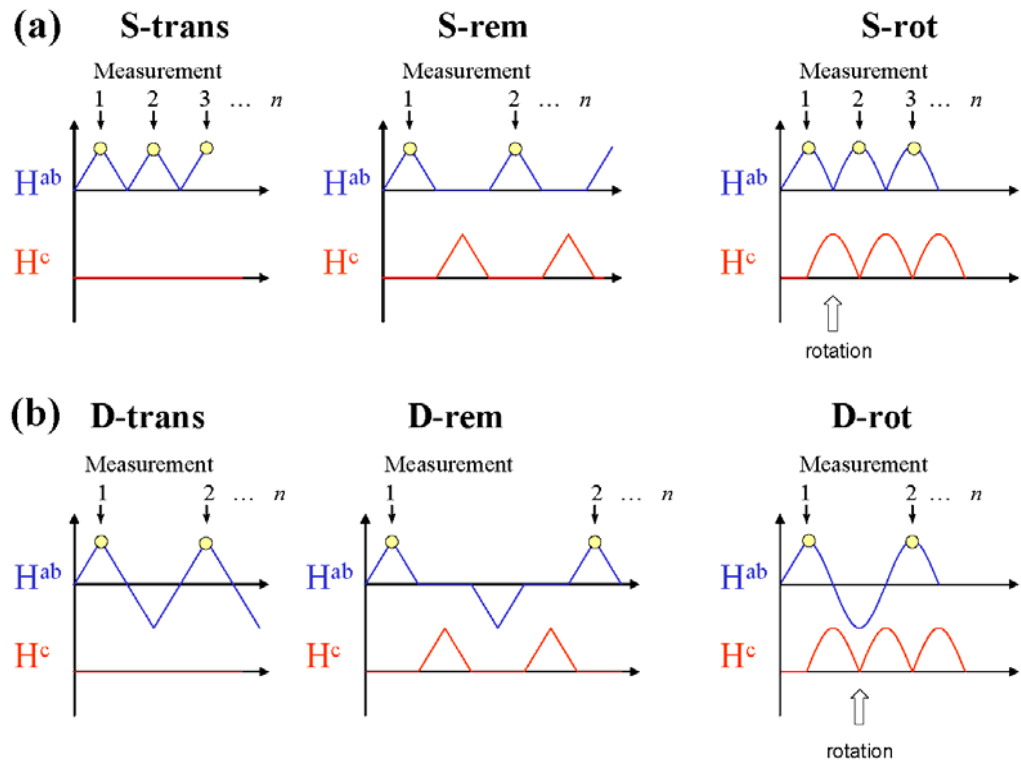

**Figure 2**

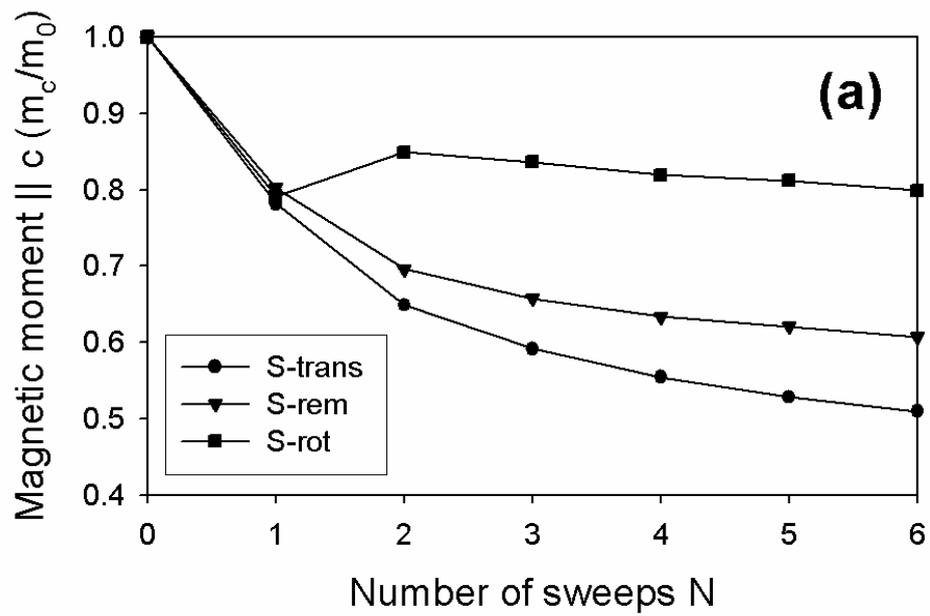

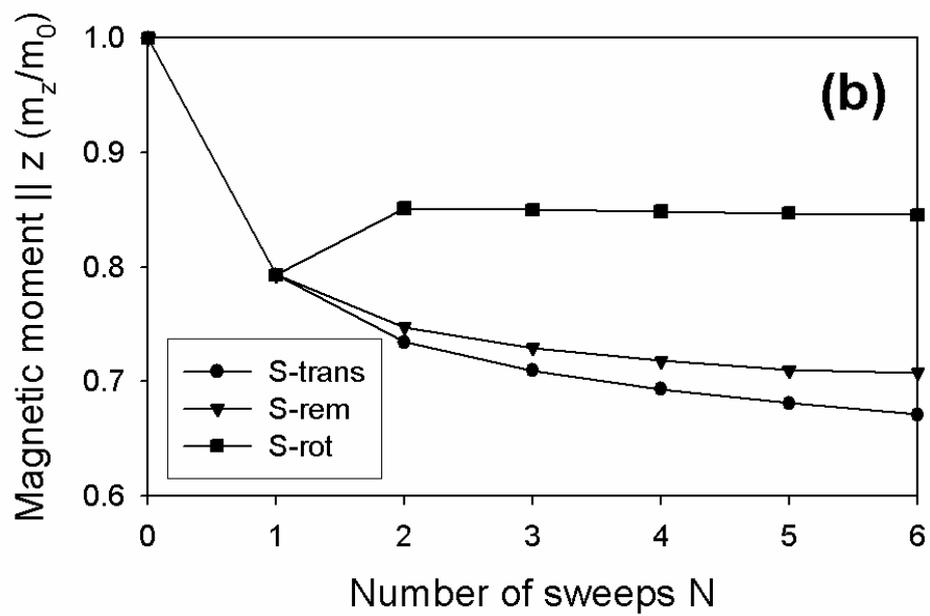

Figure 3

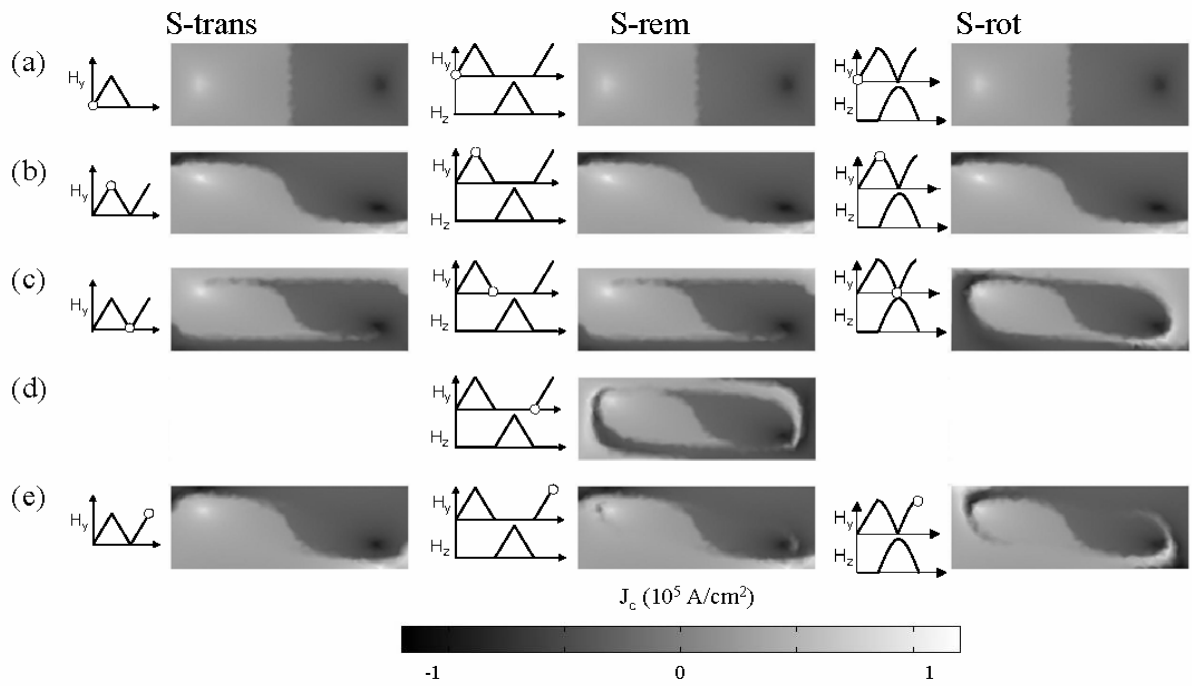

**Figure 4**

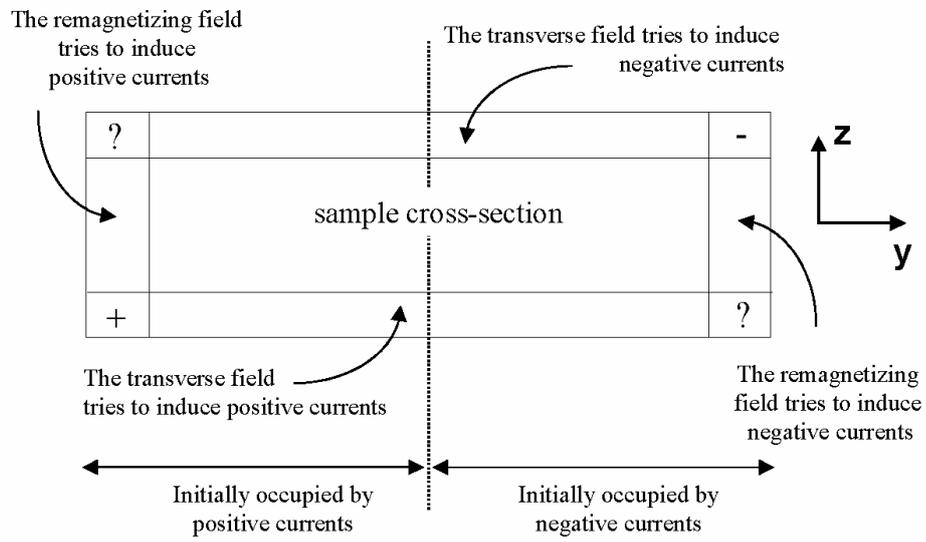

Figure 5

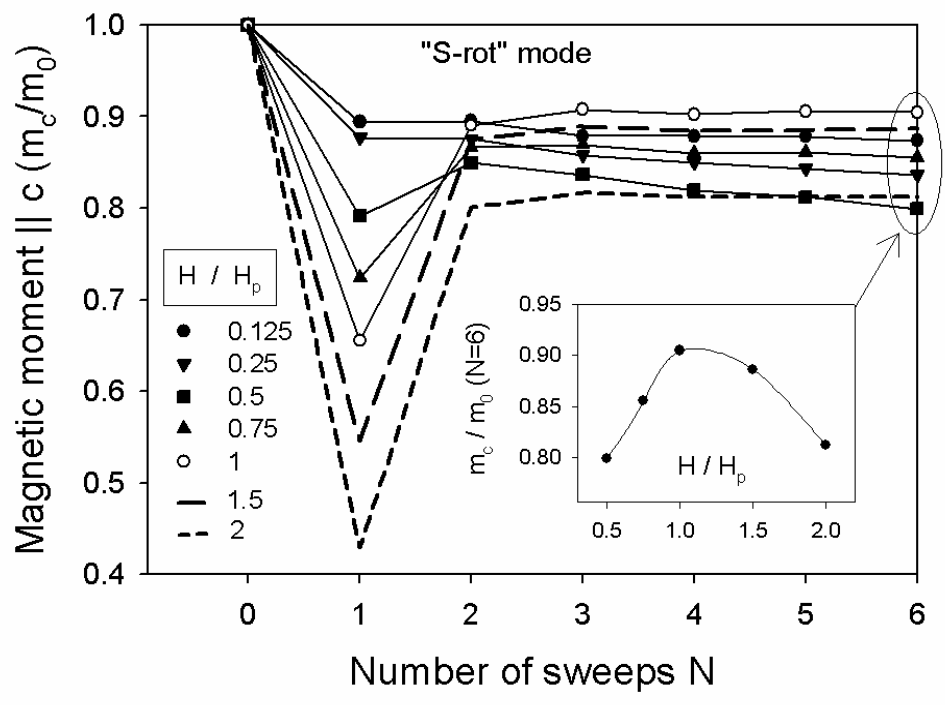

**Figure 6**

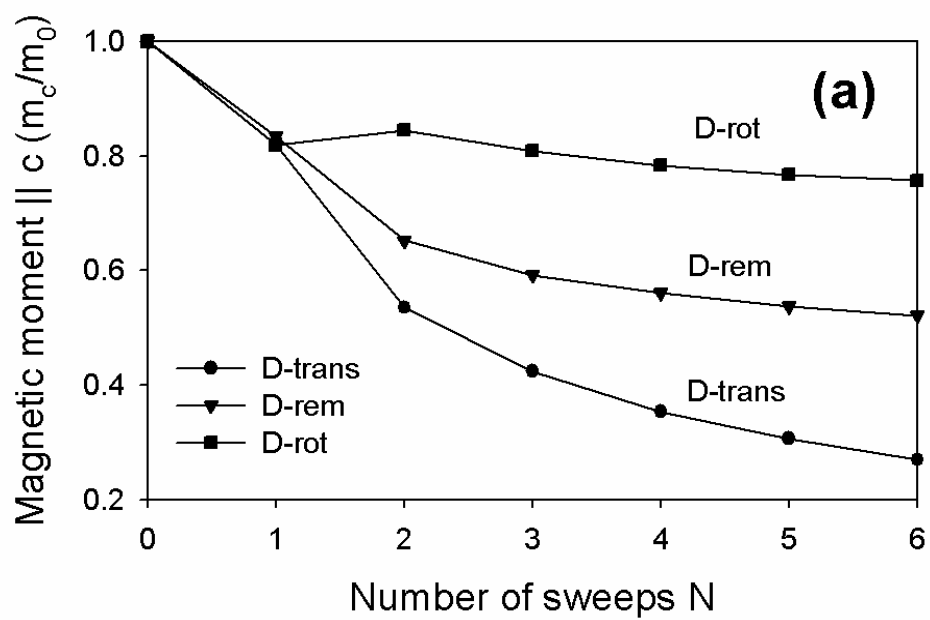
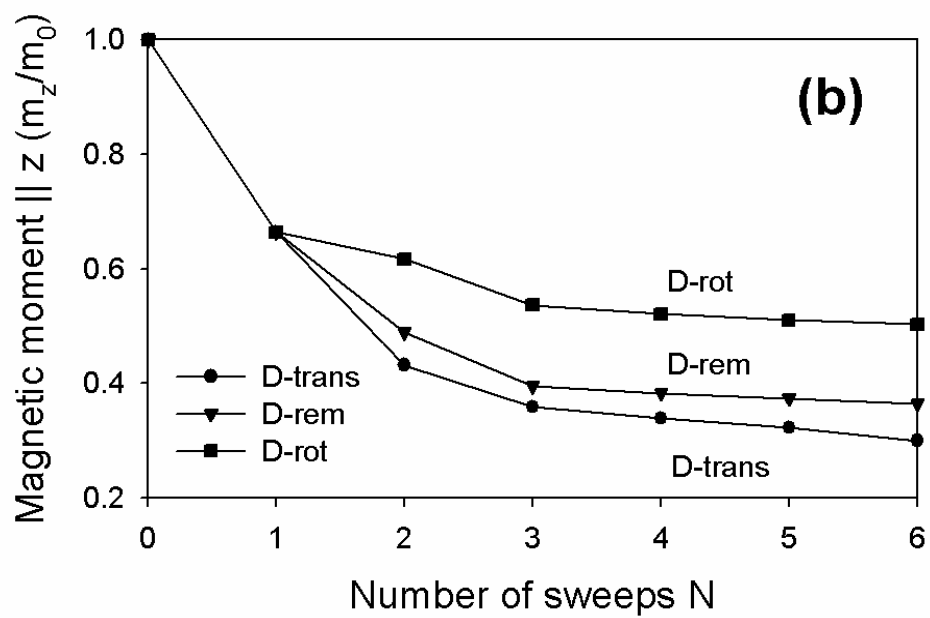

Figure 7

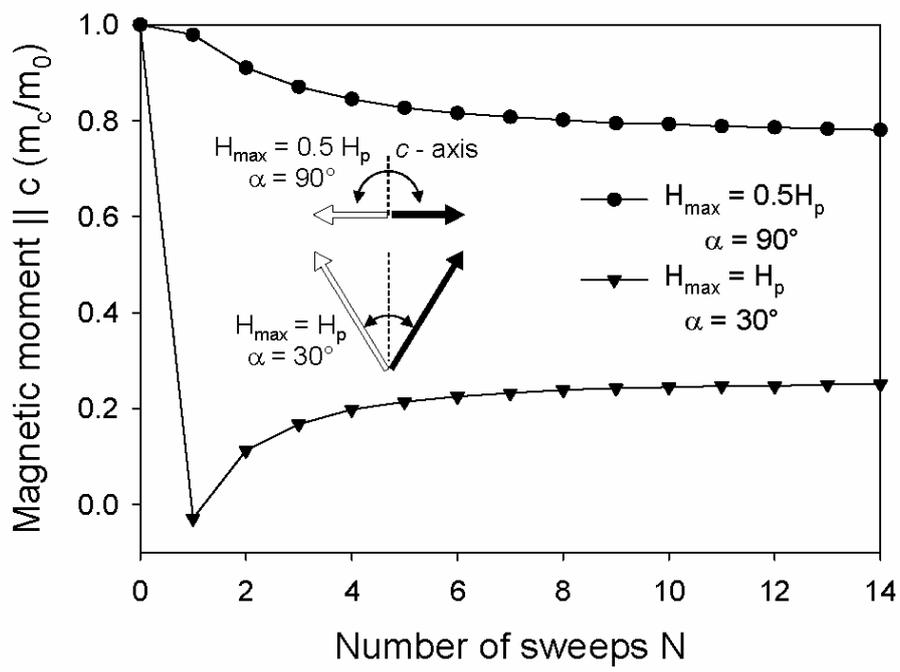

**Figure 8**

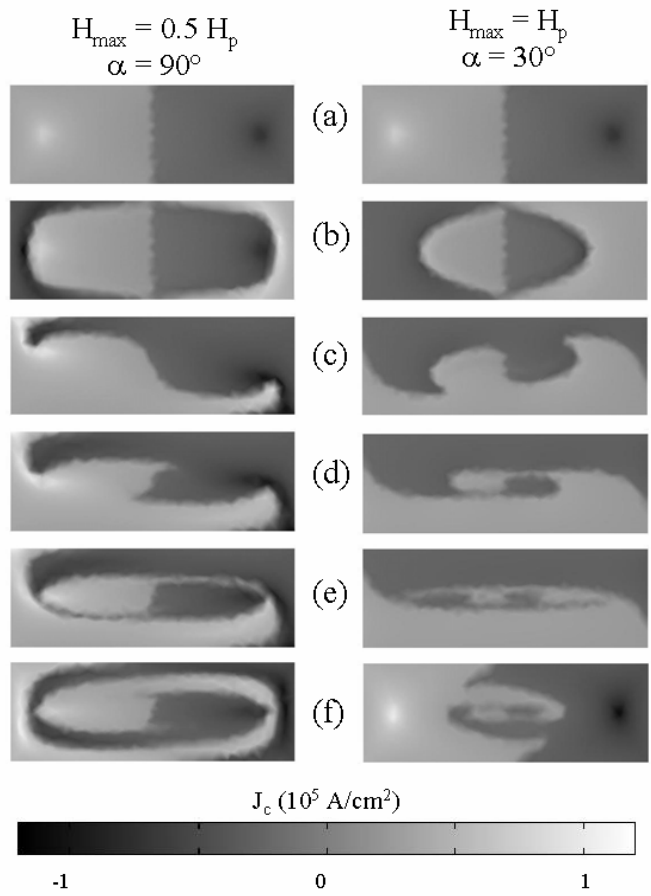

**Figure 9**